\documentclass[aps,pra,preprint,superscriptaddress,amsmath,amssymb,showpacs]{revtex4-1}
\usepackage{amssymb,bm}
\usepackage{graphicx}
\usepackage{amsmath}
\usepackage[colorlinks=true,citecolor=blue,urlcolor=blue]{hyperref}
\begin{document}

\title{Parity violation in high-energy proton-proton scattering.}
\author{A.I. Milstein}
\affiliation{Budker Institute of Nuclear Physics of SB RAS, 630090 Novosibirsk, Russia}
\affiliation{Novosibirsk State University, 630090 Novosibirsk, Russia}
\author{N.N. Nikolaev}
\affiliation{L.D. Landau Institute for Theoretical Physics RAS, 142432 Chernogolovka, Russia}
\author{S.G. Salnikov}
\affiliation{Budker Institute of Nuclear Physics of SB RAS, 630090 Novosibirsk, Russia}
\affiliation{Novosibirsk State University, 630090 Novosibirsk, Russia}
\date{\today}

\begin{abstract}
A new approach is proposed to describe the parity violation in the process of scattering of a polarized proton by a proton at high energies. It is shown that in the tree approximation the P-odd amplitude of proton-proton scattering is substantially smaller than the P-odd amplitude in proton-neutron scattering. The main source of P-odd asymmetry in proton-proton scattering is radiative corrections due to the charge-exchange strong interaction. We predict that the asymmetry in the inelastic cross section is substantially less than the asymmetry in the elastic cross section.
\end{abstract}

\maketitle

{\bf Introduction.}\\
The observation of the parity-violating dependence of the total cross section on the helicity of accelerated protons is one of the actual tasks in the program of polarization experiments at the NICA  collider \cite{NICA, Kekelidze2016vcp, Savin2014sva}. P-odd single-spin asymmetry 
$$ A = (\sigma_+ - \sigma_-)/(\sigma _++ \sigma_-)$$
 in elastic and deep inelastic scattering of longitudinally polarized electrons due to interference of weak and electromagnetic interactions has been studied both theoretically and experimentally \cite{H1998, VFVLR2000, HPSM2001, MS}. It has become one of the precision sources of information about the structure of a weak neutral current \cite { Qweak}. There is extensive theoretical \cite {BHK1974, HK1975, KF1975, FK1976, DDH1980, FS1981, BT1981, O1981, NP1982, GP1983} and experimental \cite {P1974, N1978, B1980, L1984, Y1986, E1991, B2003} literature devoted to  the effects of weak interaction on the background of strong interaction. The most accurate data on the P-odd asymmetry in total cross sections were obtained for the scattering of longitudinally polarized protons and neutrons with energies of tens of MeV and below, usually described by P-odd meson exchange potentials. The question on the asymmetry of $ A $ at high energies remains open \cite {NICA}. Assuming experiments at the NICA collider, here we discuss energies in the center-of-mass frame, much smaller than the masses of the $ W $ and $ Z $ bosons, when the microscopic weak interaction Hamiltonian can be considered as a contact current-current one. The parton language is adequate for large momentum transfers, while to describe the elastic amplitude and, accordingly, the total cross section at moderately high energies, one has to take into account the strong interaction phenomenologically and use the language of interacting baryons and mesons. As will be shown, such a description in conjunction with  the eikonal approach predicts the non-trivial isotopic properties of the P-odd nucleon interaction and the dominance of the contribution of elastic scattering to the P-odd
asymmetry in the total cross section.

 {\bf Parity violation in nucleon-nucleon scattering.} \\ 
 From an experimental point of view, the  one-spin P-odd asymmetry in the interaction of a longitudinally polarized proton with a hydrogen target is of particular interest. The microscopic weak interaction Hamiltonian $H_W^{(0)}$ has the form \cite{OKUN}:
\begin{equation}\label{H}
H_W^{(0)}=-\dfrac{G}{\sqrt{2}}[j_\mu^{ch} j_\mu^{ch}+j_\mu^{0} j_\mu^{0}]\,,
\end{equation}
 where $ G $ is the Fermi constant, $ j_\mu^{ch} $ and $ j_\mu^{0} $ are the quark charged and neutral weak currents. An account for the strong interaction in the amplitude of weak nucleon interaction involves three steps: calculating the nucleon matrix elements of weak currents taking into account the hadron  sizes, analysing the effect of radiative corrections due to strong interaction on the isotopic properties of P-odd interaction, and taking into account absorption corrections. In the nucleon basis, weak currents have the form (we omit the contribution of the so-called weak magnetism which is insignificant in our problem)
 \begin{eqnarray}\label{currents}
 &j_\mu^{ch}=\cos\theta_c\,\{\bar p\gamma_\mu(1+g\gamma_5)n+\bar n\gamma_\mu(1+g\gamma_5)p\}\,,\nonumber\\
 &j_\mu^{0}=[\bar n\gamma_\mu n+g\bar n\gamma_\mu\gamma_5n]+[(4\xi-1)\bar p\gamma_\mu p-g\bar p\gamma_\mu\gamma_5p]\,,
 \end{eqnarray}
 where $ \gamma_{\mu} $ are the Dirac matrices, $\cos\theta_c = 0.974 $, $ g = 1.26 $, and $\xi = \sin^2 \theta_W = 0.2383\pm~0.0011 $ \cite{Qweak}.
 Similar to the hadron electrodynamics, passing to a weak interaction of physical nucleons can be carried out by introducing form factors. We are only interested in the P-odd weak interaction of nucleons, $ H_{PNC} = H^0 + H^{ch} $, the isotopic structure of the matrix elements of which in the nucleon sector is substantially determined by the negligibly small factor $ | 4\xi-1 | $:
 \begin{eqnarray}\label{HPNC0}
 &H^0=G_0\,F^0_A(q_{13})F^0_V(q_{13})(\bar n_4\gamma_\mu n_2)(\bar p_3\gamma_\mu \gamma_5p_1)\,,\nonumber\\
 &H^{ch}=-G_{ch}\,F^{ch}_A(q_{14})F^{ch}_V(q_{14})[(\bar n_4\gamma_\mu\gamma_5p_1)(\bar p_3\gamma_\mu n_2)+(\bar n_4\gamma_\mu p_1)(\bar p_3\gamma_\mu\gamma_5 n_2)] +\frac{1}{N_c}\{n_4 \leftrightarrow p_3 \},\nonumber\\
 &G_0=\dfrac{gG}{\sqrt{2}}\,,\quad  G_{ch}=\dfrac{gG}{\sqrt{2}}\cos^2\theta_c\,.
 \end{eqnarray} 
 Here $ p_1 $ and $ n_2 $ are initial, and $ p_3 $ and $ n_4 $ are the final proton and neutron, respectively, $ q_i $  denote the momenta of the corresponding particles, $ q_{ij} = q_i-q_j $, $ F^{0, ch}_A(q) $ and $ F^{0, ch}_V (q) $ are the axial and vector form factors of weak currents. Recall the property of conservation of $ s $ -channel helicity: the amplitude of elastic $ np $ scattering due to $ H^0 $ is proportional to $ \lambda_1 \delta_{\lambda_3 \lambda_1} \delta_{\lambda_4 \lambda_2} $, and the amplitude of $ np $ charge exchange due to $ H^{ch} $ is proportional to $ (\lambda_1 + \lambda_2) \delta_{\lambda_3 \lambda_2} \delta_ {\lambda_4 \lambda_1} $, where $ \lambda_i = \pm 1 $ is the corresponding particle helicity. The first term in $ H^{ch} $ corresponds to the kinematics of $ np $ charge exchange, the second term, $ \propto 1 / N_c $ ($ N_c = 3 $ is the number of colours), describes elastic $ np $ scattering.
 For our purposes, it is sufficient to use the dipole approximation 
 $$
 F^{ch}_{V, A}(q) = F^{0}_{V, A}(q) =\left(1-\dfrac{q^2}{M_{V, A}^2}\right)^{- 2 } \,, $$
 with $ M_V = 0.8 \, $GeV, $ M_A = 1 \, $GeV \cite {BEM2002}.
 It is noteworthy that according to \eqref {HPNC0}, the P-odd effect is present only in the $ np $ interaction. Below we show how this property will be modified by strong interaction.
 
  {\bf Effect of radiative corrections on the isotopic asymmetry structure.}\\
  P-odd $ pp $ interaction can be induced by P-odd $ np $ interaction due to radiative corrections associated with strong interaction. Typical contributions of these radiative corrections are shown in Fig. 1, where Fig. 1a corresponds to the contribution of charged currents and Fig. 1b corresponds to the contribution of neutral currents. The latter contribution has a meaning of the renormalization of the neutral current and should not be taken into account separately. 
  \begin{figure}[h]
 	\centering
	\includegraphics[width=0.65\linewidth]{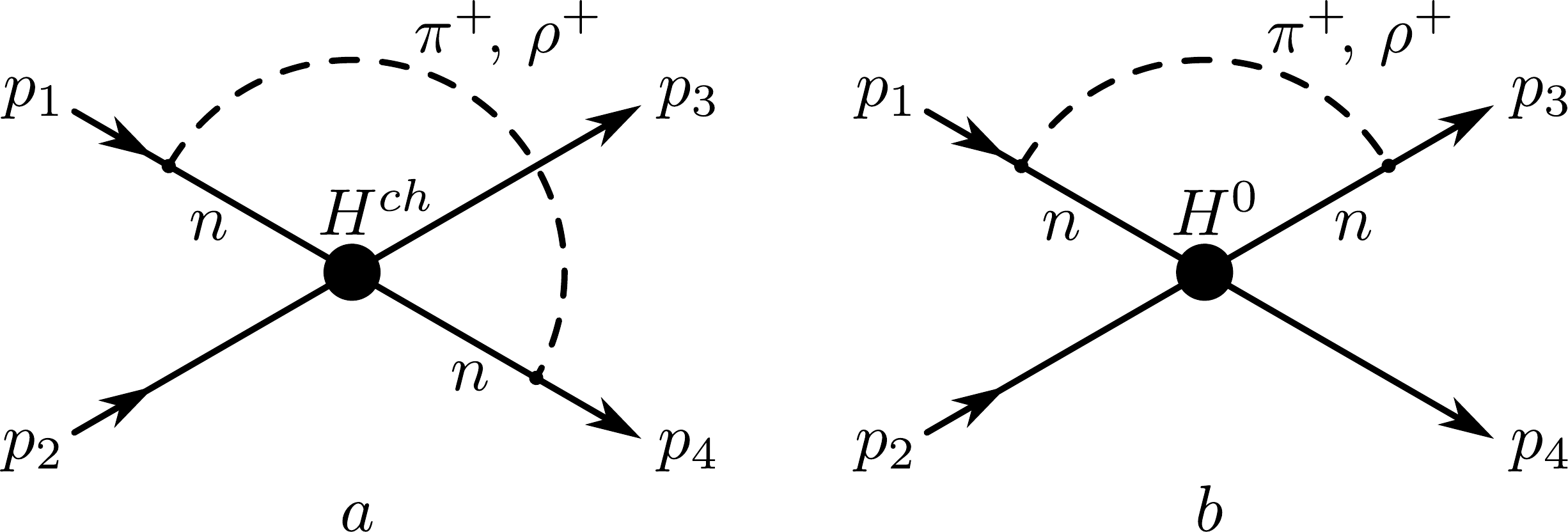}
 	\caption{Typical contributions to the parity-violating amplitude of $ pp $ scattering. Diagrams (a) and (b) correspond to the Hamiltonians $ H^{ch} $ and $ H^{0} $ in Eq.~\eqref {HPNC0}.}
 	\label{FD}
 \end{figure}
Consider the usual interaction of $ \rho $ mesons with a vector nucleon current
  $$
  L_\rho=-g_\rho\bar N\gamma_\mu{\bm \tau}\cdot{\bm \rho}_\mu N\,,
 $$
where $g_\rho^2/4\pi=0.55\pm 0.06$  \cite{EW}, ${\bm \tau}$ are Pauli isospin matrices, ${\bm \rho}_\mu$ is the $\rho$ -meson field, $ N $ is the Dirac spinor for the isodoublet of  nucleons. We are interested in $ pp $ scattering within the diffraction cone.
Taking the loop integral over the 4-momentum $ k_\mu $ of  $\rho$ mesons  using the Sudakov technique, we obtain in the leading  logarithmic approximation
\begin{eqnarray}\label{Hspiral}
 &&T_W^{(\rho)}(\bm q_\perp)=\dfrac{2}{\pi^3}G_{ch}g_\rho^2 R_\rho(\bm q_\perp)\ln\left(\dfrac{s}{M_V^2}\right)\lambda_1\,\delta_{\lambda_2\lambda_1}\delta_{\lambda_3\lambda_1}\delta_{\lambda_4\lambda_2}\, . 
 \nonumber\\
 &&R_\rho(\bm q_\perp)=\int \dfrac{F^2(k_\perp)\,d^2k_\perp}{(\bm k_\perp-\bm q_\perp)^2+m_\rho^2}\,.
 \end{eqnarray}
  In this model estimate of the function $ R_\rho (\bm q_\perp) $, for simplicity, we set $ F_A^ {ch} (q) = F_V^{ch}(q) \equiv F(q) $ and omitted the possible form factors at the $ \rho NN $ vertex.   The only important thing is that the integration converges in the region of restricted transverse momenta $ \bm k_\perp $ of the $ \rho $ meson. The logarithmic factor $\ln(s/M_V^ 2) $ arises from integration over the longitudinal momentum of the $ \rho $ meson. A similar contribution from the exchange by a charged pion is suppressed ($ \propto 1 / s $) due to the zero spin of the pion. Thus,  radiative corrections due to the strong interaction has changed the isotopic structure of the P-odd Hamiltonian. Strictly speaking, we cannot insist on the logarithmic amplification in the formula (\ref {Hspiral}), since it arises from the region of a large virtuality in  the nucleon propagator. With all reservations, the answer obtained at the hadronic level (\ref {Hspiral}) gives a reasonable scale of P-odd asymmetry for polarized protons accelerated in NICA with energies of several GeV interacting with the fixed hydrogen target.
 
 {\bf Asymmetry in the scattering cross section.}\\
  An important property of strong interaction at the discussed energies  is the conservation of $ s $-channel helicity in nucleon-nucleon elastic scattering inside the diffraction cone. Consequently, the further  discussion of the interference of amplitudes of strong and P-odd weak interactions can be performed in terms of scalar amplitudes. In the differential cross section for elastic scattering, 
  $ d\sigma/d^2q_\perp=| T(\bm q_\perp)|^2/16\pi^2$, the matrix element $T$ can be written as \begin{eqnarray}\label{Mtot}
 &&T(\bm q_\perp)=T_s(\bm q_\perp)+T_W(\bm q_\perp)+ T_{int}(\bm q_\perp)\,,\nonumber\\
 &&  T_{int}(\bm q_\perp)=-\dfrac{i}{2} \int \dfrac{d^2q'_\perp}{(2\pi)^2}\,T_s(\bm q'_\perp) T_W(\bm q_\perp-\bm q'_\perp)\,.
 \end{eqnarray}
Here $T_s(\bm q_\perp)$ is the amplitude of strong (and electromagnetic) interaction,
 $T_W(\bm q_\perp)$ is the amplitude of weak interaction with account for the radiative corrections to  P-odd Hamiltonian, $T_{int}(\bm q_\perp)$ is the so-called absorption correction to weak amplitude due to strong interaction. The representation \eqref{Mtot} can be  easily derived in eikonal formalism. 
 
   Since $T_W(\bm q_\perp)$ is a real quantity, it follows from the optical theorem that the P-odd contribution of $\delta\sigma_{tot}$ to the total cross section is solely  related to the absorption correction
    \begin{eqnarray}
    \delta\sigma_{tot}=\int \dfrac{d^2q_\perp}{8\pi^2}\mbox{Re}\Big\{T_s^*(\bm q_\perp)T_W(\bm q_\perp)\Big\}\,.                
    \end{eqnarray}
   On the other hand, the P-odd contribution of $\delta\sigma_{el}$ to the elastic cross section is
   \begin{eqnarray}
   \delta\sigma_{el}=\int \dfrac{d^2q_\perp}{8\pi^2}\mbox{Re}\Big\{T_s^*(\bm q_\perp)\Big[ T_W(\bm q_\perp)+ T_{int}(\bm q_\perp)\Big]\Big\}\,.                
   \end{eqnarray}
  Here, the first contribution exactly coincides with $\delta\sigma_{tot}$. Then the second contribution coming from $T_{int}(\bm q_\perp)$ corresponds with a minus sign to the P-odd contribution of $\delta\sigma_{in}$ to the inelastic cross section. We show that there are reasons for the smallness of $\delta\sigma_{in}$ compared to $\delta\sigma_{tot}$ and $\delta\sigma_{el}$.

   The main contribution to the elastic cross section is determined by the region of the diffraction cone with standard parameterization of the elastic scattering amplitude 
   $$
   T_s(\bm q_\perp)=-(\epsilon+i)\exp(- q_\perp^2\beta^2/2)\,\sigma_{tot}\,,$$
   where $\sigma_{tot}$ and $\beta^2$ are the energy-dependent total cross section and slope of the diffraction cone, respectively. The phase of the elastic scattering amplitude determined by the parameter $ \epsilon $ also depends on the energy. It is important that, within the diffraction cone, $ \epsilon $ has a relatively weak dependence on the  momentum transfer 
   \cite {SH2010, FO2013}. In this case, the phase $T_{int} (\bm q_\perp) $
   differs from the phase of the elastic scattering amplitude by $ \pi / 2 $, which means the vanishing of the contribution $\delta\sigma_{in}$ to the inelastic cross section. This is important when planning experiments, since the extraction of elastic scattering allows one to enhance the observed P-odd asymmetry.
   
    Now we present numerical estimates of the expected P-odd effect. For the energies discussed, the dependence of $T_{W}(\bm q_\perp)$  on the  momentum transfer, determined by the form factors, 
    is close to this dependence for the amplitude $T_s(\bm q_\perp)$. Then, for the example of proton-neutron scattering due to neutral currents, we obtain
      \begin{align}\label{aspn}
      &{\cal A}_{tot}^{pn}=-\dfrac{4 \sqrt{2}\, g G\epsilon_{pn}\sigma_{el}^{pn}}{(1+\epsilon_{pn}^2)(\sigma_{tot}^{pn})^2}\,.
      \end{align}
    Here we assume that $ \epsilon $ is  independent of the  momentum transfer and use the relation between the elastic cross section and the slope of the diffraction cone:
    $$\sigma_{el}=\dfrac{(1+\epsilon^2)\sigma_{tot}^2}{16\pi\beta^2}\,.$$
    The contribution of charged currents to $ A_{pn} $ differs from the contribution of neutral currents by the factor $ \cos^ 2 \theta_c /N_c $, it is determined by the last term in \eqref {HPNC0} for $ H^{ch}$. For $ s  = 13 \,\mbox{GeV}^2$  we have: $ \epsilon_ {pn} = - 0.5 $, $ \sigma_{tot}^{pn} \sim 50 \, $ mbn, $ \sigma_{el}^{pn} \sim 15 \, $ mbn \cite {R2003}. The asymmetry in the total $ np $ scattering cross section  is positive and, with account for both of the above contributions,  equals to $ {\cal A}_{tot}^{pn}\sim 10^{- 7} $.
    
    We estimate the asymmetry in $ pp $ scattering using the amplitude (\ref {Hspiral}):
   \begin{align}\label{as}
   &{\cal A}_{tot}^{pp} \sim -G_{ch}g_\rho^2 \dfrac{0.1\,\epsilon}{\pi^3 \beta^2}\ln\left(\dfrac{s}{M_V^2}\right)\,.
   \end{align}
    The coefficient 0.1 in this formula is the rounded result of numerical integration with the model formula (\ref {Hspiral}) for $ R_\rho (q) $. With the same $ s = 13 \, \mbox{GeV}^2 $  we have: $ \beta^ 2 \sim 9\,\mbox{GeV}^{-2} \, $, $ \epsilon = - 0.5 $, $ \sigma_{tot} \sim 50 \, $ mbn, $ \sigma_{tot} / \sigma_{el} \sim 3.5 \, $. The asymmetry in the total cross section is positive and equals to $ {\cal A}_{tot}^{pp} \sim 0.4 \cdot 10^{- 7} $. We conclude that the P-odd asymmetry in the total $ pp $ scattering cross section is several times smaller than the asymmetry in the total $ np $ scattering cross section.
   
{\bf Conclusion.}\\
We have proposed a new approach to describe  parity violation in  scattering of polarized protons by unpolarized protons at moderately high energies of the NICA collider. It is shown that radiative  corrections due to the charge exchange strong interaction generate a non-zero P-odd proton-proton scattering amplitude. According to our estimates, the P-odd asymmetry in proton-proton scattering is noticeably smaller than the asymmetry in proton-neutron scattering. Therefore, the search for P-odd asymmetry in the scattering of polarized protons should be carried out on neutron-containing nuclear targets. It follows from the analysis of absorption corrections  that in inelastic scattering the P-odd asymmetry is suppressed, so that it is preferable in the experiment to extract an elastic channel  in which the asymmetry is enhanced: $ {\cal A}_{el}/{\cal A}_ {tot} \approx \sigma_{tot} / \sigma_{el} $. In the NICA energy region, this gain is noticeable one: $ \sigma_{tot} / \sigma_{el} \sim 3.5\, $.

{\bf Acknowledgements.}\\
 We are grateful to I.A. Koop and Yu.M. Shatunov for useful  discussions.
This work is supported  by RFBR Grant No. 18-02-40092 MEGA.

\end{document}